\documentclass[aps,apl,twocolumn,showpacs]{revtex4}
\usepackage{graphicx}
\usepackage{hyperref}
\begin{document}

\title{Large local Hall effect in pin-hole dominated multigraphene spin-valves.}

\author{P. K. Muduli, J. Barzola-Quiquia, S. Dusari, A. Ballestar, F. Bern, W. B\"{o}hlmann, P. Esquinazi}
\affiliation{Division of Superconductivity and Magnetism, Institut
f$\ddot{u}$r Experimentelle Physik II, Universit$\ddot{a}$t
Leipzig, Linn$\acute{e}$stra${\ss}$e 5, D-04103 Leipzig, Germany}
\email{muduli@physik.uni-leipzig.de}
\date{\today}
\begin{abstract}
We report local and non-local measurements in pin-hole dominated
mesoscopic  multigraphene spin-valves. Local spin-valve
measurements show spurious switching behavior in resistance during
magnetic field sweeping similar to the signal observed due to
spin-injection into multigraphene. The switching behavior has been
explained in terms of local Hall effect due to thickness
irregularity of the tunnel barrier. Local Hall effect appears due
to large local magnetostatic field produced at the roughness in
the AlO$_x$ tunnel barrier. The effect of this local Hall effect
is found to reduce as temperature is increased above 75 K. The
strong local Hall effect hinders spin-injection into multigraphene
resulting in no spin signal in non-local measurements.
\end{abstract}
\pacs{XX-XX, XX-XX-XX}
%\keywords{}
\maketitle
% start main body here................
\clearpage
\section{Introduction}
Graphene and few-layer graphene has shown to be ideal material for
spintronics devices because of suppressed hyperfine interaction
and weak spin-orbit coupling\cite{hernando}. Gate tunable
conductivity along with low carrier density ($n$ $<$10$^{12}$
cm$^{-2}$) and high mobility ($\mu$ $\sim$10$^{4}$
cm$^{2}$V$^{-1}$s$^{-1}$) gives opportunity to fabricate
ultra-fast spintronic devices with this material\cite{geim}.
Efficient spin-polarized carrier injection into graphene is one of
the essential requirements for such devices. Significant
improvement has been achieved in this direction through successful
demonstration of spin injection into single and few-layer graphene
(also called multigraphene) at room temperature by different
groups \cite{tombrosnature,hill,cho,nishioka,wang,jo,goto,tb}.
Long spin diffusion length $\lambda_s$ $\sim$2 $\mu$m and spin
life time $\tau_{sp}$ $\sim$50-200 ps have been reported in single
layer graphene through spin injection
experiments\cite{tombrosnature}. Much longer spin life time
$\tau_{sp}$ $\sim$2 ns has been observed in bilayer
graphene\cite{han,yang}. Few-layer graphene is now believed to be
a more appropriate candidate for spin injection compared to single
layer graphene because of the screening of scattering potentials
from substrate in the former\cite{massen, yang,garcia}. However,
experimental values of $\lambda_s$ and $\tau_{sp}$ are still an
order of magnitude shorter than what is expected
theoretically\cite{hernando}. Therefore, an efficient engineering
of spin-valve devices is required to realize the full potential of
multigraphene spintronics devices.

Spin-valve devices for spin injection show high to low resistance
switching, which depends on the relative magnetization of the two
ferromagnetic electrodes. Spin polarized carriers are injected
from one ferromagnet (injector) and the probability of these
carriers to be collected at the other ferromagnet (detector)
depends on the magnetization of the latter. Therefore, quality of
interface between the ferromagnet and graphene may significantly
affect the spin-polarized carrier injection into graphene. A
proper understanding of the micromagnetic phenomena occurring at
the interface is essential for the optimization of such devices.
In order to avoid conventional Schottky barrier contact due to the
large conductivity mismatch between graphene and the metallic
ferromagnet, a tunneling approach has been
suggested\cite{hanprl,jozsa}. However, conventional tunnel
barriers suffer from complications like pin-holes and nonuniform
barrier thickness. These defects in the barrier give additional
features by contributions from anisotropic magnetoresistance (AMR)
and local Hall effect that obscure spin injection
signal\cite{hong}. In particular, stray magnetostatic fields
produced at the interface due to the ferromagnet on top of it may
have significant impact. Recently, spin precession and inverted
Hanle effect has been reported in Si spin injection devices which
arise from local magnetostatic fields at tunnel barrier
roughness\cite{das}.

In this work, we demonstrate large local Hall effects due to stray
magnetostatic fields at the tunnel barrier roughness in
multigraphene spin-valves. We observe resistance switching
behavior due to this local Hall effect. We found that spin
injection is strongly hindered by the presence of these stray
magnetostatic fields. No resistance switching was observed in
non-local measurements due to this.
\begin{figure}[!h]
\begin{center}
%\vskip -1.5cm
\abovecaptionskip -10cm
\includegraphics [width=4 cm]{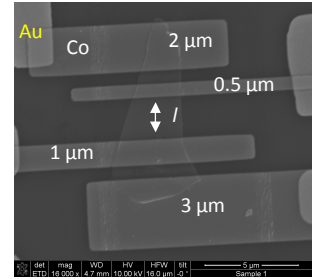}
\end{center}
\caption{\label{fig1} Scanning electron microscope (SEM) image of
one of the spin valve devices fabricated by electron beam
lithography.}
\end{figure}

\section{Fabrication and experimental details}
Multigraphene samples were prepared on Si/Si$_3$N$_4$ substrate by
rubbing small flake of highly oriented pyrolytic graphite (HOPG)
from Advanced Ceramics (grade ZYA, rocking curve 0.4$^o$).
Thickness of the multigraphene flakes were measured using atomic
force microscope (AFM). For device fabrication $\sim$30 nm thick
and $\sim$10 $\mu$m long multigraphene samples were chosen.
Spin-valve devices were fabricated by conventional electron beam
lithography method. Thermally evaporated Al on top of
multigraphene was oxidized to form AlO$_x$ tunnel barrier. The
thickness of the AlO$_x$ tunnel barrier was found out to be
$\sim$2 nm from AFM measurements. For spin polarized carrier
injection into multigraphene $\sim$50 nm thick Co was thermally
evaporated on a pre-patterned structure of PMMA prepared by
electron beam lithography. Ferromagnetic Co lines of different
width (as shown in Fig. 1) were patterned using this technique.
Different width of Co was used to ensure different coercive fields
between electrodes. The distance between two inner Co electrode
($\emph{l}$) was varied from 1-3 $\mu$m for different spin-valve
devices. A thin layer of Pt was immediately evaporated on top of
Co to prevent it from oxidation. The Co lines were further
contacted to larger contact pads through gold lines.
Magnetoresistance measurements were done in a closed cycle
refrigerator using an AC resistance bridge. For spin valve
measurements magnetic field was applied in the plane of the film
along the long axis of the Co line using an electromagnet with
rotation option.

\section{Results and discussion}
Temperature dependence of resistance of the spin-valve device
showed a semiconducting behavior as typically observed for
multigraphene samples reported elsewhere\cite{arndt,garcianjp}.
Current-Voltage (IV) characteristics were measured on as
fabricated samples at room temperature and 15 K. Spin-valve
devices showed slightly nonlinear IV with few kOhm resistance due
to the presence of AlO$_x$ tunnel barrier. However, with repeated
IV measurement with maximum current up to 10 $\mu$A the devices
showed linear IV with resistance decreasing down to $\sim$435 Ohm
at 15 K. The rapid decrease of resistance with linear IV can be
understood by considering formation of pin-holes during the
measurement process. The resistance of the spin-valves was
measured with an AC resistance bridge with 1.5 $\mu$A current
limit range.
\begin{figure}[!h]
\begin{center}
%\vskip -1.5cm
\abovecaptionskip -10cm
\includegraphics [width=7 cm]{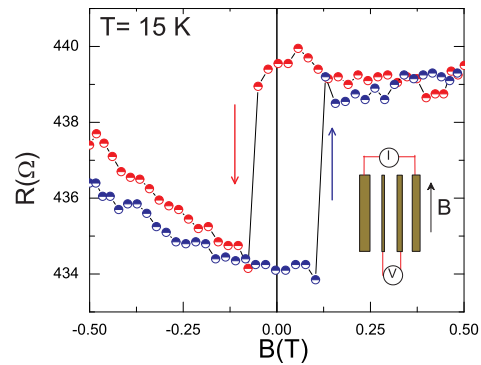}
\end{center}
\caption{\label{fig1} Local spin-valve measurements at T = 15 K in
the configuration as shown in inset. High to low and low to high
resistance switch can be seen  when the magnetic field is swept
from positive to negative and vice versa for magnetic field
applied along the long axis of the Co electrode(as shown by
arrow).}
\end{figure}
Fig. 2 shows the field dependence of the resistance of one of the
spin-valve device measured at 15 K. The measurement configuration
called local configuration is shown in the inset of Fig. 2.
Magnetic field ($B$) is applied along the long axis of the Co
line. Resistance switching can be seen at a magnetic field $B
\sim$0.1 T. The resistance switches from higher resistance to a
lower resistance while field is swept from +0.5 T to -0.5 T. A
reverse case happens while field is swept from -0.5 T to +0.5 T.
The direction of field sweeping is shown by arrows in the figure.

\begin{figure}[!h]
\begin{center}
%\vskip -1.5cm
\abovecaptionskip -10cm
\includegraphics [width=7 cm]{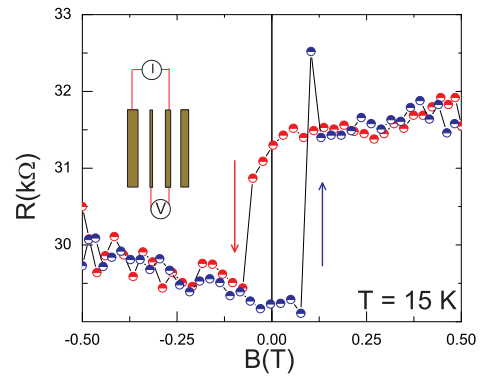}
\end{center}
\caption{\label{fig1} Local spin-valve measurements in 3-terminal
local configuration as shown in inset. Magnetic field is applied
along the long axis of the Co electrodes.}
\end{figure}
Fig. 3 shows a similar resistance switching behavior for a
3-terminal local configuration as shown in the inset. Our
resistance switching behavior is significantly different from
resistance switching seen in local measurements in graphene spin
injection devices reported in literature\cite{tombrosnature,han}.
In spin injection devices step-like switching is seen for parallel
and antiparallel orientation of the ferromagnetic electrodes. High
resistance state is achieved for antiparallel configuration of
inner two electrodes and a low resistance for parallel
configuration. Our switching behavior is quite unusual and can be
misunderstood as signature of spin injection.
\begin{figure}[h]
\begin{center}
%\vskip -1.5cm
\abovecaptionskip -2cm
\includegraphics [width=9 cm]{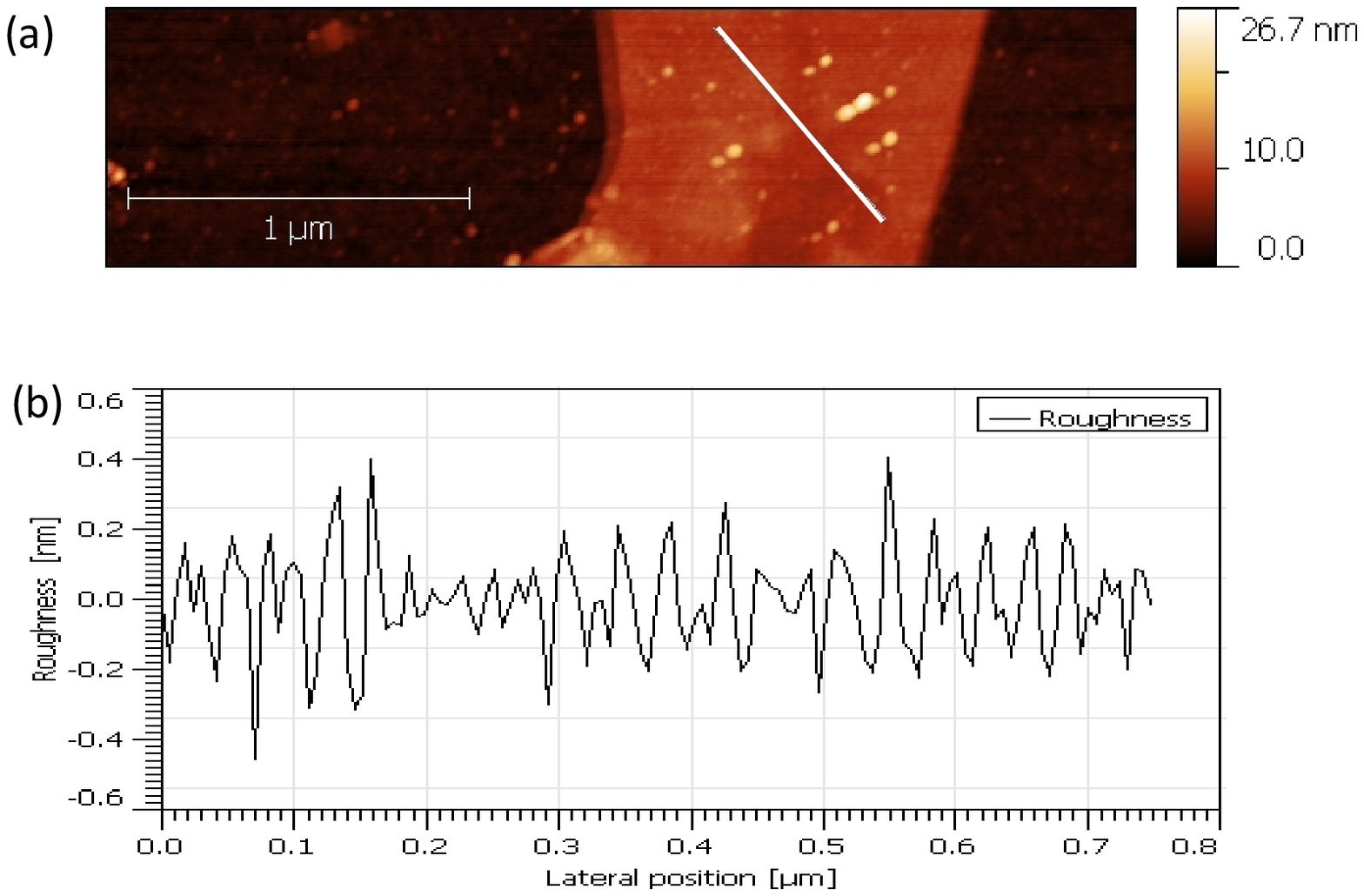}
\end{center}
\caption{\label{fig1} (a) AFM image of multigraphene with AlO$_x$
on top of it. (b) Roughness profile along the line in (a)}
\end{figure}
The observed switching behavior can be understood by considering
the presence of different stray magnetostatic fields in the
spin-valve device. Magnetostatic fields in a spin-valve device
appears mainly from two sources: (1) Fringe magnetostatic fields
at the edges of the ferromagnetic electrodes \cite{monzon}and (2)
Magnetostatic fields near the multigraphene and ferromagnet (Co)
interface due to finite roughness of the tunnel
barrier\cite{das}.With the magnetic field applied along the long
axis of the Co electrode fringe fields from the edges can be
ignored in our spin-valves as the long edges of the Co electrode
lie far from the multigraphene (see Fig. 1). Most of the
magnetostatic fields in our device mainly originate from the
AlO$_x$ roughness. As we observe a linear IV characteristics, the
AlO$_x$ tunnel barrier in our spin-valve is highly nonuniform with
a lot of pin-holes.  So when the ferromagnetic Co electrode is
magnetized due to roughness of the oxide barrier, a local
magnetostatic field with a strong normal component $B_z$ can be
generated. In a very simplified situation with one dimensional
periodic roughness along $x$ axis the magnetostatic field
generated from a ferromagnet with magnetization pointing along the
$x$ direction can be written as\cite{das,nogaret},
\begin{equation}
\begin{array}{l}
 B_x (x,z) = \mu _o M_s \left( {\frac{h}{2}} \right)\sum\limits_{n = 1}^\infty  {q_n F(} q_n )e^{ - q_n z} \sin (q_n x - \pi /2) \\
 B_y (x,z) = 0 \\
 B_z (x,z) = \mu _o M_s \left( {\frac{h}{2}} \right)\sum\limits_{n = 1}^\infty  {q_n F(} q_n )e^{ - q_n z} \cos (q_n x - \pi /2) \\
 \end{array}
\end{equation}.
Where $ q_n  = 2\pi n/\lambda$ and
\begin{equation}
F(q_n ) = \frac{{\sin (q_n \lambda /4)}}{{(q_n \lambda
/4)}}\frac{{\sinh (q_n h/2)}}{{(q_n h/2)}}
\end{equation}

\begin{figure}[h]
\begin{center}
%\vskip -1.5cm
\abovecaptionskip -2cm
\includegraphics [width=6 cm]{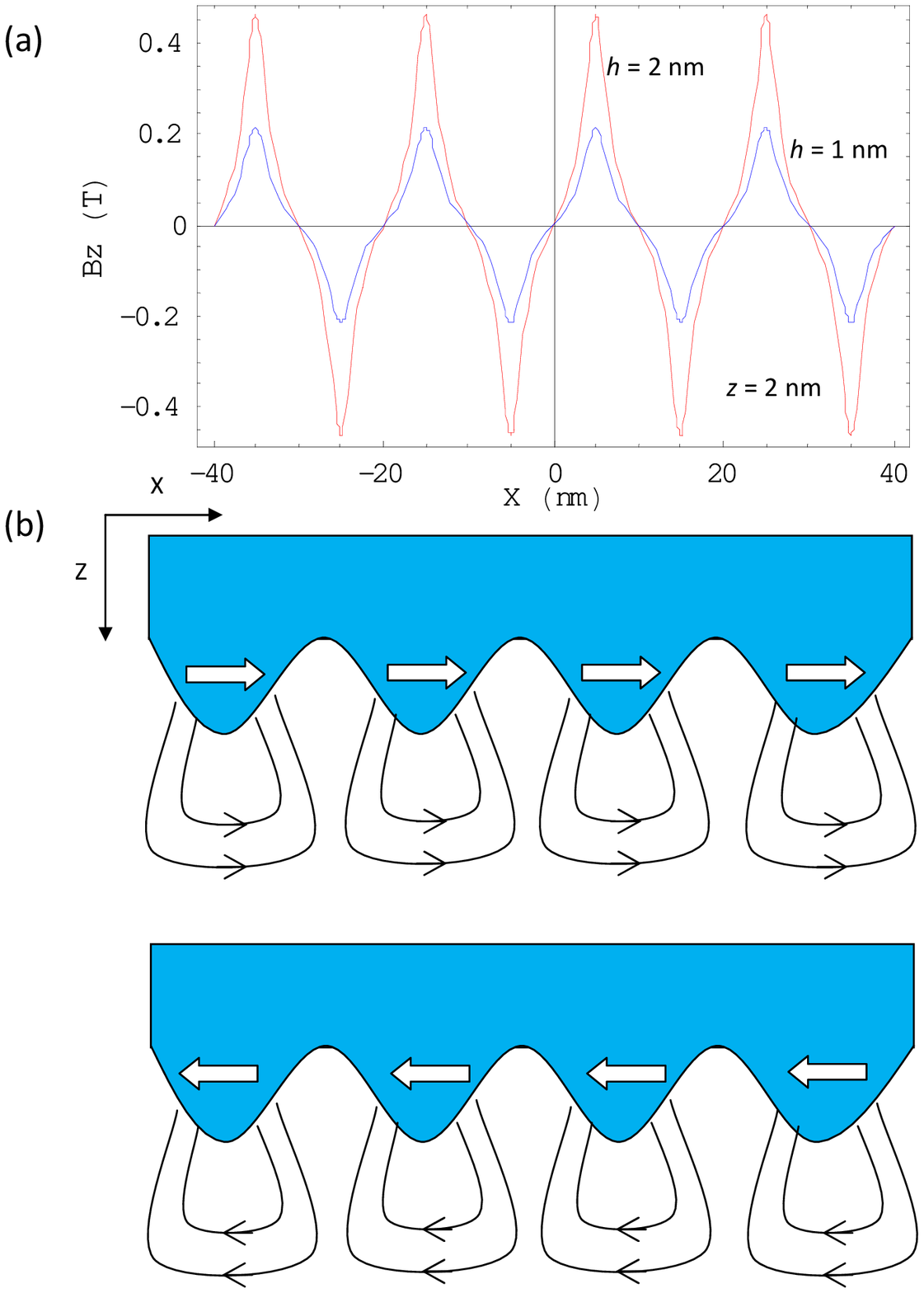}
\end{center}
\caption{\label{fig1}(a) The out of plane component of magnetic
field $B_z$ calculated from Eq. (1) for two different roughness
amplitude $h$= 1 and 2 nm. Here  $z$ = 2 nm and  $\lambda \sim $
20 nm .(b) Sketch of magnetostatic field from a sinusoidal
roughness profile for two different direction of magnetization f
the Co electrode. Magnetic poles developed at the interface is
shown by arrows. }
\end{figure}
Here $h$ is the peak-to-peak roughness height and $\lambda$ the
period of the roughness profile of the barrier. To get a detailed
idea about the roughness of AlO$_x$ tunnel barrier we performed
AFM measurements of our spin valve as shown in Fig. 4(a).
Roughness profile of AlO$_x$ on top of multigraphene along the
marked line in Fig. 4(a) is shown in Fig. 4(b). Note that in our
spin-valves multigrphene is covered with AlO$_x$ all over it
before patterning the Co electrodes. Clearly, the roughness
amplitude in the AlO$_x$ layer is $h\sim$1 nm. However,
considering the linear IV characteristics and the possibility to
have pin-holes we can assume a maximum roughness amplitude $h\sim$
2 nm i.e, of the order of the thickness of the AlO$_x$ barrier.
For ferromagnetic Co with $ \mu _0 M_s = 1.82$ T, and assuming $
\lambda \sim$20 nm the out of plane component of magnetic field
($B_z$) calculated using Eq. (1) at $z$ = 2 nm is shown in Fig. 5
(a) for $h$ = 1 and 2 nm, respectively. The local magnetostatic
fields can be as large as $\sim$0.45 T. A pictorial representation
of the stray magnetostatic fields for a sinusoidal roughness of
the tunnel barrier is shown in Fig. 5 (b) with magnetization
pointing in two opposite direction. Therefore, when an external
field $\sim$0.1 T is applied to saturate the Co electrode, the
multigraphene with a tunnel barrier roughness $\sim$2 nm can
experience much higher field of the order of 0.45  T in local
areas. We believe these stray magnetostatic fields will provide
additional voltage contribution $ V^{lh}$ to the measured
longitudinal voltage $ V^m _{xx}$. The value of $ V^{lh}$ will
depend on the actual distribution of the stray magnetostatic
fields, and therefore on the roughness. However, in a more
qualitative way one can say that in a mesoscopic device with
irregular geometry voltage between any two points can be written
as a linear combination of longitudinal ($ V_{xx}$) and Hall ($
V_{xy}$) voltages  . Therefore, one can write, $ V^{lh} =aV _{xx}+
bV _{xy} $, where $a$ and $b$ are sample dependent constants. When
the magnetization of Co electrode changes direction the
perpendicular component of the magnetostatic field $B_z$, changes
sign resulting in change in the sign of Hall voltage $V_{xy}$.
Therefore, the measured voltage $ V^m _{xx}$, for $B_z$, pointing
upward, can be written as,$ V^m _{xx}  = V_{xx} + V^{lh}=
(a+1)V_{xx}  + bV_{xy}$,  and in this case one gets a high
resistance state. Similarly, for $B_z$ pointing downwards one can
write, $ V^m _{xx} = (a+1)V_{xx}  - bV_{xy}$ and thus a low
resistance state is achieved. The switching from the high
resistance to low resistance state occurs at a magnetic field when
magnetization of the Co electrode changes, i.e, at the coercive
field of the inner Co electrode. Although one expects two
resistance switching corresponding to the magnetization switching
of the two inner Co electrodes, we found only one resistance
switching probably because the narrower Co electrode produces too
small resistance change to be observable within our experimental
resolution. We also measured our spin-valves at different
temperatures and we found that the resistance change produced by
this local Hall effect becomes too small and in the range of
experimental noise above 75 K. Field dependence of resistance
measured  at T = 50, 60, 75 and 100 K in the local configuration
(see Fig. 2) is shown in Fig. 6.

Another possible source of switching in resistance in spin-valves
can come from anisotropic magnetoresistance (AMR) of the  Co
electrodes\cite{filip}. However, we could not find any sharp
resistance switching in spin-valves with  Co electrodes without
any AlO$_x$ tunnel barrier\cite{heuse}. Therefore, we believe the
sharp resistance switching in our spin-valves is mainly due to
local Hall effect.

\begin{figure}[h]
\begin{center}
%\vskip -1.5cm
\abovecaptionskip -2cm
\includegraphics [width=8 cm]{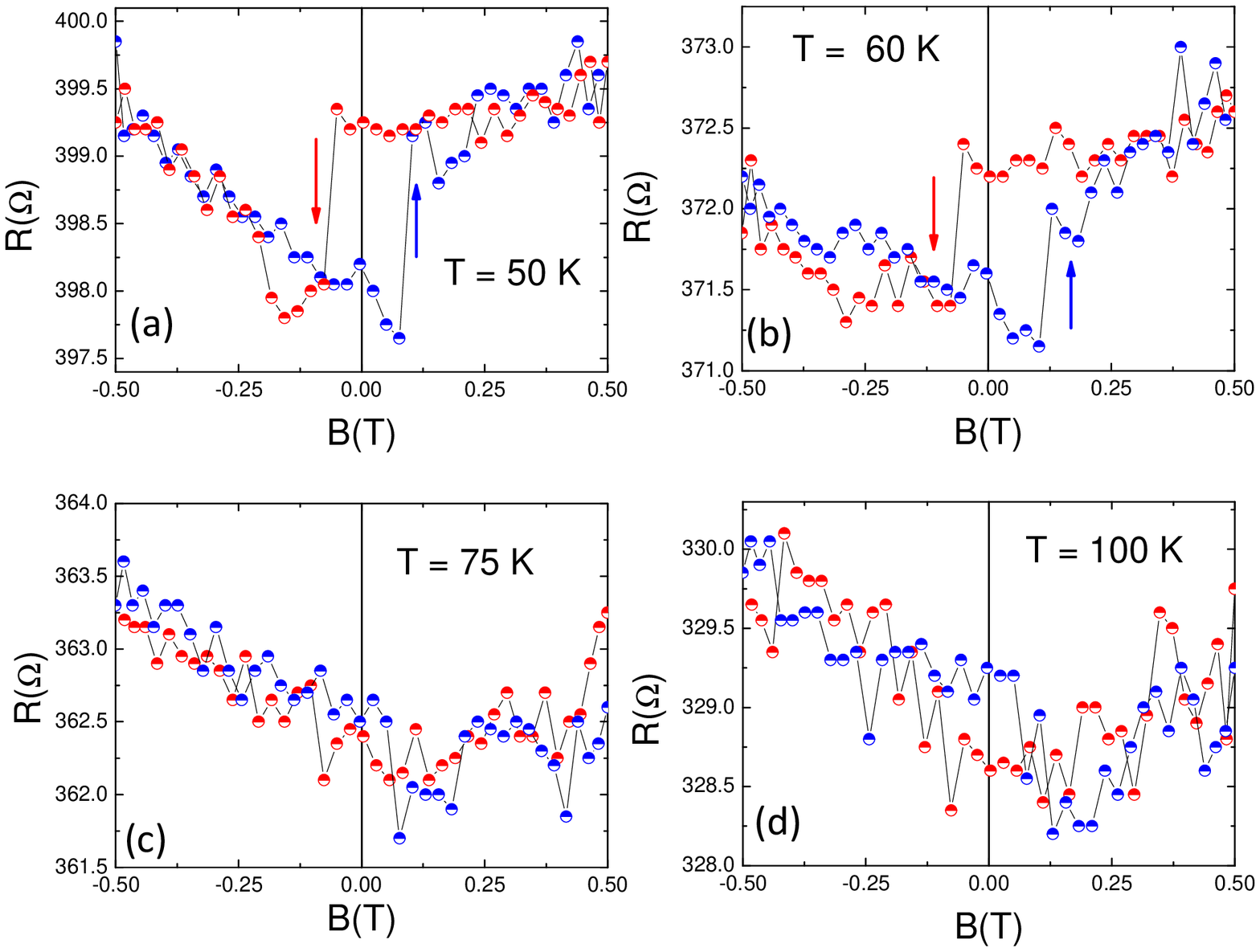}
\end{center}
\caption{\label{fig1}Local spin-valve measurements at (a) T = 50
K, (b) T = 60 K, (c) T = 75 K and (d) T= 100 K in the
configuration as shown in inset of Fig. 2.Absence of resistance
switching above 75 K shows reduced effect of local Hall effect.}
\end{figure}
Clear signature of spin injection into multigraphene is usually
seen in non-local measurements in a configuration as shown in the
inset of Fig. 7. In this configuration, electrical charge current
path is completely separated from spin current path so that
signals only due to spin current can be observed. In our devices,
we could not find any switching in resistance in non-local
measurement done at 15 K with voltage pads placed 1.5 $\mu$m away
from the current path. We believe the local Hall effect due to
tunnel barrier roughness significantly suppress spin injection
into multigraphene. Although, there might be additional reasons
for the absence of non-local signal in our spin-valves, we believe
that local magnetostatic fields at the tunnel barrier roughness
definitely plays  a role in spin-injection. Therefore, for a clear
observation of resistance changes due to spin injection into
multigraphene full coverage of spin-transport channel with a
smooth tunnel barrier is essential.

\begin{figure}[h]
\begin{center}
%\vskip -1.5cm
\abovecaptionskip -2cm
\includegraphics [width=7 cm]{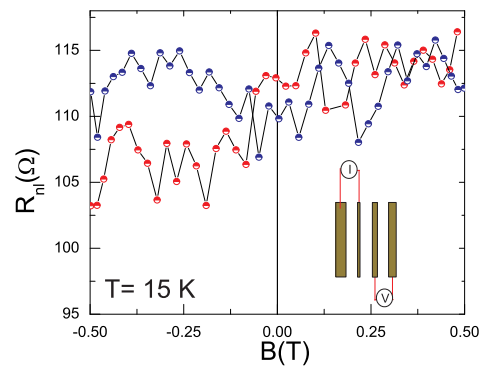}
\end{center}
\caption{\label{fig1}Non-local spin-valve measurements done at 15
K in the configuration as shown in the inset. Absence of
resistance switching indicates suppressed spin injection.}
\end{figure}
\section{Conclusions}

In summary, we have fabricated submicron spin-valve devices with
few-layer graphene of different thickness. Our spin-valves showed
resistance switching behavior at low temperatures in local
measurement configuration. However, in non-local configuration we
could not see any switching indicating reduced spin injection. The
switching behavior in local measurement can be qualitatively
understood in terms of  large local Hall effect. The local Hall
effect appears due to magnetostatic fields present in the device.
We estimated that for a roughness $\sim$ 2 nm the local
magnetostatic fields can be as large as 0.45 T. Local Hall effect
can result in spurious resistance switchings, which can be
confused with resistance switching due to spin injection into spin
valve systems. Pin-hole free smooth barriers along with narrow
ferromagnetic electrodes are prerequisite for spin-injection into
multigraphene.

\begin{acknowledgements}
This research has been supported by grants from Deutsche
Forschungsgemeinschaft under Contract No. DFG ES 86/16-1.SD and AB
acknowledge financial support by Graduate School of Natural
Sciences BuildMoNa of the University of Leipzig.
\end{acknowledgements}

\end{document}